 \documentclass[12pt,preprint]{aastex}

 \usepackage{emulateapj5}

\slugcomment{to appear in the Astrophysical Journal}

\shorttitle{Recurrent Novae CI Aquilae}
\shortauthors{Hachisu et al.}

\begin{document}

\title{REVISED ANALYSIS OF THE SUPERSOFT X-RAY PHASE, HELIUM ENRICHMENT, AND
TURN-OFF TIME IN THE 2000 OUTBURST OF THE RECURRENT NOVA CI AQUILAE}


\author{Izumi Hachisu}
\affil{Department of Earth Science and Astronomy, 
College of Arts and Sciences, University of Tokyo,
Komaba, Meguro-ku, Tokyo 153-8902, Japan} 
\email{hachisu@chianti.c.u-tokyo.ac.jp}

\author{Mariko Kato}
\affil{Department of Astronomy, Keio University, 
Hiyoshi, Kouhoku-ku, Yokohama 223-8521, Japan} 
\email{mariko@educ.cc.keio.ac.jp}

\and

\author{Bradley E. Schaefer}
\affil{Department of Astronomy, University of Texas, Austin, TX 78712}
\email{schaefer@astro.as.utexas.edu}




\begin{abstract}
     Recurrent nova CI Aquilae has entered the final decline phase
a bit before May of 2001, about 300 days 
after the optical maximum, showing the slowest evolution
among recurrent novae.  Based on the optically thick 
wind mass-loss theory of
the thermonuclear runaway model, we have estimated the turn-off time of 
the CI Aql 2000 outburst.  It is in late March of 2001 
after a luminous supersoft
X-ray source phase lasts $\sim 150$ days (from November of 2000
until March of 2001).  We have also obtained,
by fitting our theoretical light curves 
with the 1917 and 2000 outbursts, 
the white dwarf (WD) mass to be $M_{\rm WD}= 1.2 \pm 0.05 ~M_\sun$, 
the helium enrichment of the envelope is He/H$\sim 0.5$ by number,
the mass of the hydrogen-rich envelope on the WD at the optical maximum  
is $\Delta M_{\rm max} \sim 8.0 \times 10^{-6} M_\sun$, 
and the average mass accretion rate is
$\dot M_{\rm acc} \sim 1.0 \times 10^{-7} M_\sun$ yr$^{-1}$ 
during the quiescent phase between the 1917 and 2000 outbursts. 
Using these obtained values, we have consistently reproduced 
the light curve in quiescence as well as of the two outbursts.
We have also discussed the possibility whether or not CI Aql 
will explode as a Type Ia supernova in a future.
\end{abstract}




\section{INTRODUCTION AND SUMMARY}
     The second recorded nova outburst of CI Aquilae 
was discovered on 2000 April 28 UT by \citet{tak00}, 83 years 
after the first recorded outburst in 1917 \citep{rei25,wil00db}. 
CI Aql now becomes a member of the recurrent nova class. 
About 300 days after the optical maximum,
\citet{hac01ka} estimated various physical parameters 
of the CI Aql system from the light curve fitting and
elucidated its nature.  They derived
the white dwarf (WD) mass to be $M_{\rm WD}= 1.2 \pm 0.05 ~M_\sun$, 
the helium enrichment of the white dwarf envelope 
to be He/H$\sim 0.25$ by number, and
the mass of the hydrogen-rich envelope 
on the white dwarf at the optical
maximum to be $\Delta M_{\rm max} \sim 5.8 \times 10^{-6} M_\sun$. 
This envelope mass indicates an average mass accretion rate of
$\dot M_{\rm acc} \sim 0.7 \times 10^{-7} M_\sun$ yr$^{-1}$ 
during the quiescent phase between the 1917 and 2000 outbursts. 
They finally predicted the turn-off time of 2001 August and,
therefore, that the luminous supersoft X-ray source phase 
lasts until August of 2001.  
\par
     After \citet{hac01ka} has been published, 
some new observational indications appeared concerning the
turn-off time of the 2000 outburst.
The supersoft X-ray fluxes were too weak 
during June and August of 2001 \citep{gre02}, 
which indicates that the steady hydrogen shell-burning had 
already vanished.
\citet{sch01b} made optical photometry of CI Aql in August of 2001 
and concluded that CI Aql was $V \sim 15.3$ and almost 
near its quiescent phase \citep{men95}.  
These results contradict the theoretical prediction 
that, if the hydrogen content of the white dwarf envelope
is $X=0.5$ by weight, the steady hydrogen shell-burning
lasts until August of 2001 \citep{hac01ka}.  
\citet{sch01c} found on 12 Harvard College Observatory archival 
photographs the 1941 outburst, the brightness of which is 
$B \sim 13.8$ on May 2 of 1941, and suggested that the 1917 discovery
could indicate a recurrence timescale of about 20 yrs, with missed
outbursts around 1960 and 1980.
In this paper, therefore, we reanalyze the light curves of CI Aql,
especially reproducing the orbital light curves newly obtained 
in April/May and August of 2001 \citep[see also][]{sch01a, sch01b}.
\par
     Additionally, Honeycutt (2002, private communication) has 
revised Mennickent \& Honeycutt's (1995) photometric data of
CI Aql in quiescence.
The revised data show about 0.62 mag down in the $V$-light curve so
that the quiescent level is as low as $V \sim 16.1$, 0.8 mag
fainter than August of 2001.  This value is well consistent 
with Szkody's (1994) $V \sim 16.2$ in quiescence 
\citep[see also Table 2 of ][]{men95}.
In this sense, CI Aql did not reach its quiescent level but still
showed an activity even in August of 2001.  Therefore, taking
these new things into account, we have fully reconsidered the CI Aql
model.
\par
     To summarize the main revised results, \\
{\bf 1.} The estimated white dwarf mass is the same as 
the previous result, i.e., $M_{\rm WD}= 1.2 \pm 0.05 ~M_\sun$.  \\ 
{\bf 2.} Helium enrichment of the white dwarf envelope is estimated
to be He/H$\sim 0.5$ by number, much larger than the previous value. \\
{\bf 3.} Optically thick winds stopped in early November of 2000 and
the estimated turn-off time is not in August of 2001 but
earlier than May of 2001, i.e., March or April of 2001. 
The luminous supersoft X-ray source phase lasted from November of 2000
until March or April of 2001.  \\
{\bf 4.} The mass of the hydrogen-rich envelope on the white dwarf 
at the optical maximum is more massive, i.e.,
$\Delta M_{\rm max} \sim 8.0 \times 10^{-6} M_\sun$ than the previous
result. 
This envelope mass indicates an average mass accretion rate of
$\dot M_{\rm acc} \sim 1.0 \times 10^{-7} M_\sun$ yr$^{-1}$ 
between the 1917 and 2000 outbursts. \\
{\bf 5.} These satisfy the conditions of Type Ia supernova explosion
if the white dwarf consists of carbon and oxygen.
\par
In \S2, we describe our basic model of CI Aql in the late phase of
the 2000 outburst and in quiescence.  
In \S3, we have theoretically modeled orbital light curves
of CI Aql in the late phase of the 2000 outburst.  
In \S4, the 1917 outburst is numerically re-analyzed, and
we re-determine the mass of the white dwarf. 
In \S5, the 2000 outburst is also numerically re-analyzed 
and the duration of the strong, optically thick wind 
and the supersoft X-ray source phase are again estimated.
Discussion follows in \S6.

\section{BASIC MODEL IN LATE PHASE}
\begin{figure}
\plotone{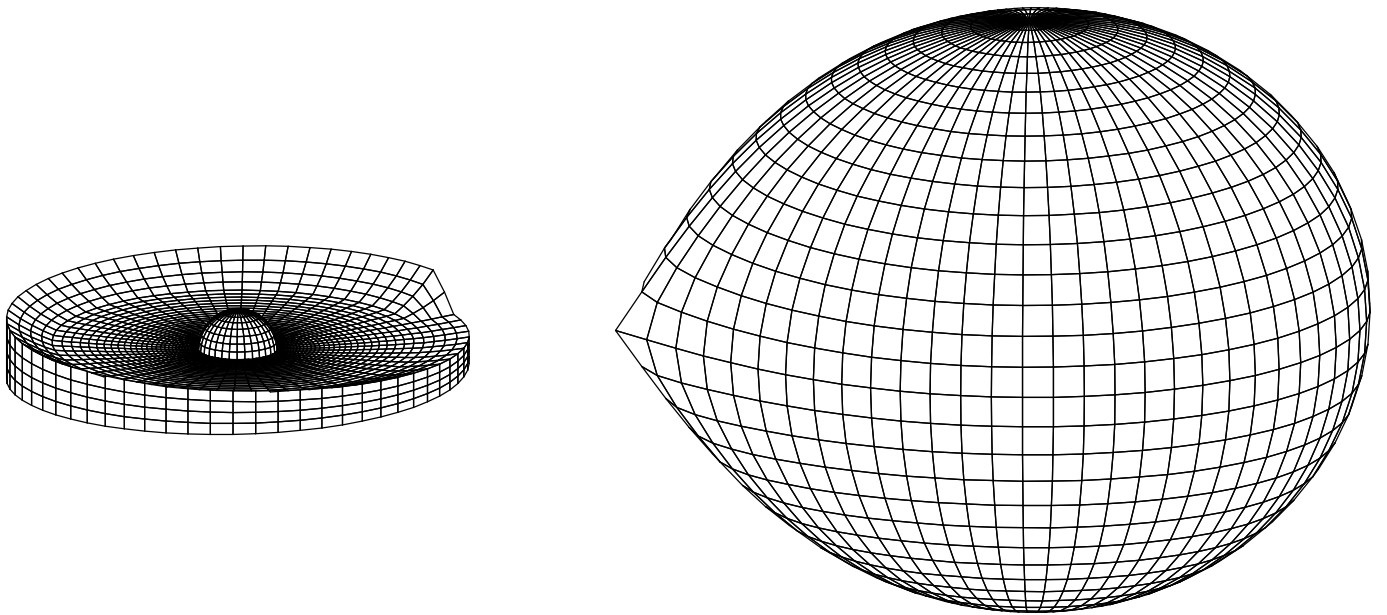}
\caption{
Configuration of our CI Aql model in the late phases of the 2000 outburst.
The cool component ({\it right}) is a slightly evolved 
MS companion ($1.5 M_\odot$) filling up its inner critical Roche lobe.  
The north and south polar areas of the cool component are 
irradiated by the hot component ($1.2~M_\odot$ white dwarf, {\it left}).
Here, we assume a non-axisymmetric accretion disk
in order to reproduce asymmetric light curves 
as shown in Figure \ref{vmag_quiescence_late_m15}.
The separation is $a= 4.25 R_\odot$; 
the effective radii of the inner critical Roche lobes are
$R_1^*= 1.53 R_\odot$, and $R_2^*= R_2= 1.69 R_\odot$, 
for the primary white dwarf and the secondary main-sequence 
companion, respectively.  The radius of the white dwarf in the cooling
phase is as small as $R_{\rm WD}= 0.0068 R_\sun$, about 1/200 times the
Roche lobe size.  The WD radius is exaggerated in this figure
so that it may be seen easily. 
\label{ciaql_late_config}}
\end{figure}

     Our binary model is illustrated 
in Figure \ref{ciaql_late_config}, which consists 
of a main-sequence star (MS) filling
its Roche lobe, a white dwarf (WD) photosphere, 
and an accretion disk.  A circular orbit is assumed.
We also assume that the surfaces of the white dwarf,
the main-sequence companion, and the accretion disk 
emit photons as a blackbody at a local 
temperature which varies with position.
For the basic structure of the accretion disk,
we assume an axi-symmetric structure with
the size and thickness of
\begin{equation}
R_{\rm disk} = \alpha R_1^*,
\label{accretion-disk-size}
\end{equation}
and
\begin{equation}
h = \beta R_{\rm disk} \left({{\varpi} 
\over {R_{\rm disk}}} \right)^\nu,
\label{flaring-up-disk}
\end{equation}
where $R_{\rm disk}$ is the outer edge of the accretion disk,
$R_1^*$ is the effective radius of the inner critical Roche lobe 
for the white dwarf component,
$h$ is the height of the surface from the equatorial plane, and
$\varpi$ is the distance on the equatorial plane 
from the center of the white dwarf.  Here, we adopt $\varpi$-square law 
($\nu=2$) 
to mimic the effect of flaring-up at the rim of the accretion disk
\citep*[e.g.,][]{sch97}.  These two parameters of $\alpha$ and $\beta$
are determined by light curve fittings.

\placefigure{ciaql_late_config}

\par
     As the August 2001 orbital light curves 
show non axi-symmetric shape, we introduce an asymmetric 
configuration of the accretion disk as done by \citet{sch97}.
They explained an asymmetric feature in the orbital light curves 
of the LMC luminous supersoft X-ray source CAL 87.  The cause is due to
the gas stream from the companion hitting the edge of the accretion disk
to make a vertical spray when the mass
transfer rate is as high as in the luminous supersoft X-ray sources.
We have also introduced the same type of asymmetry
of the accretion disk as Schandl et al.'s, i.e.,
\begin{equation}
s = {{1-\zeta_{\rm low}} \over {2 \pi - \phi_2 + \phi_1}},
\end{equation}
\begin{equation}
\zeta_{\rm edge} = \left\{ \begin{array}{ll}
 1 - s (\phi - \phi_2), &\mbox{~for~} \phi_2 \le \phi < 2 \pi \cr
 \zeta_{\rm low}+(1-\zeta_{\rm low}){{\phi-\phi_1} \over {\phi_2-\phi_1}} 
&\mbox{~for~} \phi_1 \le \phi < \phi_2 \cr
 1 - s (2\pi - \phi_2 + \phi), &\mbox{~for~} 0 \le \phi < \phi_1 
\end{array}
\right.
\end{equation}
and
\begin{equation}
{{z} \over {h}} = \left\{ \begin{array}{ll}
\zeta_{\rm low}+(\zeta_{\rm edge}-\zeta_{\rm low})
{{\xi-f_a} \over {1-f_a}},
& \mbox{~for~} \xi \ge f_a, \cr
\zeta_{\rm low}
& \mbox{~for~} \xi < f_a, \cr
\end{array}
\right.
\end{equation}
where 
\begin{equation}
\xi= {{\varpi} \over {R_{\rm disk}}},
\end{equation}
and $\zeta_{\rm low}$ is a parameter specifying the degree of asymmetry,
$\phi_1$ is the starting angle of the vertical spray, 
$\phi_2$ is the angle where the vertical spray 
reaches its maximum height, $f_a$ is the starting radius ratio 
from where the vertical spray is prominent, 
$z$ is the height of the disk surface from the equatorial plane
for the asymmetric case.
Here, we assume that the accretion disk is symmetric inside 
$\varpi < 0.8 R_{\rm disk}$ ($f_a=0.8$), and $\zeta_{\rm low}= 0.25$,
$\phi_1= (17/16) \pi$,  $\phi_2= (9/8) \pi$, unless otherwise specified.  
Such an example of the accretion disk is shown 
in Figure \ref{ciaql_late_config}.
\par
     The accretion luminosity of the white dwarf \citep*[e.g.,][]{sta88},
\begin{equation}
L_{\rm WD} =  L_{\rm WD, 0} + 
{1 \over 2} {{G M_{\rm WD} \dot M_{\rm acc}} \over {R_{\rm WD}}},
\label{accretion-luminosity}
\end{equation}
is introduced, where $L_{\rm WD}$ is the total luminosity of the 
white dwarf, $L_{\rm WD,0}$ is the intrinsic luminosity
of the white dwarf, and $R_{\rm WD}$ is the radius of the white dwarf
(e.g., $R_{\rm WD}= 0.0068 R_\sun$ for $1.2 M_\sun$ WD).
The accretion luminosity is as large as 
300 $L_\sun$ for a mass accretion rate of
$\dot M_{\rm acc} \sim 1 \times 10^{-7} M_\sun$ yr$^{-1}$
onto a $1.2 M_\sun$ WD, although it is negligibly smaller than
the luminosity of steady hydrogen shell-burning during the outburst,
$L_{\rm WD} \sim 26,000$---$54,000 ~L_\sun$, for a $1.2~M_\sun$ WD.
\par
     The viscous luminosity of the accretion disk is also introduced.
Then, the original disk surface temperature is given by 
\begin{equation}
\sigma T_{\rm disk, org}^4 = {{3 G M_{\rm WD} \dot M_{\rm acc}} 
\over {8 \pi \varpi^3}}.
\end{equation}
The temperature of the accretion disk surface is as high as
$T_{\rm disk, org} \sim 8500$ K at the radius 
$\varpi = 1~R_\sun$ for $M_{\rm WD}= 1.2 ~M_\odot$ 
and $\dot M_{\rm acc}= 1 \times 10^{-7} M_\sun$ yr$^{-1}$.
\par
     The accretion disk is heated by radiation from a luminous
white dwarf.  Such an irradiation effect is calculated as
\begin{equation}
\sigma T_{\rm disk}^4 \approx \eta_{\rm DK}
{{L_{\rm WD}} \over {4 \pi \varpi^2}}\cos\theta
+ {{3 G M_{\rm WD} \dot M_{\rm acc}} 
\over {8 \pi \varpi^3}},
\label{irradiation_disk_angle}
\end{equation}
where $\eta_{\rm DK}$ is the efficiency of the irradiation
and $\cos\theta$ is the inclination angle of the surface 
against light ray from the white dwarf \citep[e.g.,][]{sch97}.  
We assume $\eta_{\rm DK}=0.5$, unless otherwise
specified.  This means that 50\% of the absorbed radiation energy
is reemitted from the disk surface while the residual 
50\% is converted to thermal energy 
that eventually heats the white dwarf. 
The irradiation luminosity (or temperature) is much higher than 
the accretion luminosity (or temperature) during the hydrogen shell
burning, i.e., $T_{\rm disk} \sim 35,000$ K at the radius 
$\varpi = 1~R_\sun$ for $M_{\rm WD}= 1.2 ~M_\odot$, $\cos \theta= 0.1$,
and $L_{\rm WD}= 26,000~L_\sun$. 
\par
     The outer rim of the accretion disk is not irradiated 
by the white dwarf photosphere so that the temperature of the disk rim 
is simply assumed to be uniform, $T_{\rm disk, rim}$, which is one of
the parameter determined by fitting.
\par
     The surface temperature of the irradiated main-sequence (MS)
companion is roughly estimated by
\begin{equation}
\sigma T_{\rm MS}^4 \approx \eta_{\rm MS}
{{L_{\rm WD}} \over {4 \pi r^2}}\cos\theta 
+ \sigma T_{\rm MS, org}^4,
\label{irradiation_MS_angle}
\end{equation}
where $\eta_{\rm MS}$ is the efficiency of the irradiation,
$r$ is the distance from the white dwarf, 
and $\cos\theta$ is the inclination angle of the surface.
The irradiated temperature is as high as $T_{\rm MS} \sim 26,000$ K 
for $\eta_{\rm MS}=0.5$, $\cos\theta=0.5$, $M_{\rm WD}= 1.2 ~M_\sun$ 
($L_{\rm WD}= 26,000~L_\sun$), and $r \sim 4 R_\sun$.  
Therefore, the effect of the irradiation becomes very important 
when the irradiated hemisphere faces toward the Earth.
\par
     The temperature of each surface patch is determined by calculating
all contributions from every patch of the white dwarf surface
\citep[see][for more detail]{hac01kb}.  
We have divided the white dwarf surface into 16 pieces 
($\Delta \theta = \pi/16$) in the latitudinal angle and 
into 32 pieces ($\Delta \phi = 2 \pi/32$) in the longitudinal angle 
while the surface of the main-sequence companion is split into 
$32 \times 64$ pieces, respectively.  
     The surface of the accretion disk is divided into 32 pieces
logarithmically evenly in the radial direction and into 64 pieces 
evenly in the azimuthal angle.
The outer edge of the accretion disk is also divided into 64 pieces
in the azimuthal direction and 4 pieces in the vertical direction
by rectangles.  
When the photosphere of the white dwarf becomes very small, e.g.,
$R_{\rm disk}/R_{\rm ph,WD} > 10$, we attribute the first 16 meshes 
to the outer region (from $\varpi=R_{\rm disk}$ to 
$\varpi=R_{\rm disk}/\sqrt{10}$) to avoid coarse meshes in the outer part, 
and then 16 meshes to the inner region (from 
$\varpi=R_{\rm disk}/\sqrt{10}$ to $\varpi=R_{\rm ph}$), 
each region of which is divided logarithmically evenly.  
The efficiencies of irradiation of the main-sequence companion
and of the accretion disk are assumed 
to be $\eta_{\rm MS}=0.5$ and $\eta_{\rm DK}=0.5$, respectively.

\section{LATE PHASE ORBITAL LIGHT CURVES}
     To detect the end of the steady hydrogen shell-burning,
we have used photometry of CI Aql in May and August of 2001 
(see Figs. \ref{vmag_quiescence_late_m10}-\ref{vmag_quiescence_late_m20}).
All photometry was with the 2.1-m telescope at McDonald Observatory with
the IGI detector through a $V$-filter.  
The detector is a standard Tektronix 1026$\times$1024
CCD chip while the filter is the usual Schott glass sandwich as
prescribed by \citet{bes90} hence resulting in a close match to the
Johnson $V$-filter.  Standard IRAF procedures and
programs were used to get differential photometry of CI Aql with respect
to several calibration stars from Henden \& Honeycutt (1995).  A total of
118 and 429 magnitudes were obtained in April/May and August 2001, with a
typical photometric uncertainty of 0.02 mag.  
The photometric data are wrapped in one orbital phase, 
with a new ephemeris obtained by \citet{mat01}, i.e.,
\begin{equation}
t(\mbox{HJD})= 2,451,701.2086 + 0.61835 \times E,
\label{new_ephemeris}
\end{equation}
at eclipse minima.  
\par
     The orbital light curve of CI Aql in quiescence were obtained 
by \citet{men95}.  Recently, Honeycutt (2002, private communication)
rectified the data.  Essential points are that their new and old 
orbital light curves are nearly indistinguishable in shape 
but the new data are about 0.6 mag fainter than the old ones.  
We have shown these two light curves in  
Figures \ref{vmag_quiescence_late_m10}-\ref{vmag_quiescence_late_m20}.
\par
     To fit our theoretical light curves with the orbital photometry,
we have calculated $V$ light curves by changing the parameters of
$\alpha=0.5$---2.0 by 0.1 step, $\beta=0.05$---0.50 by 0.05 step,
$T_{\rm MS,org}= 5000$---10000 K by 100 K step,
$T_{\rm disk, rim}= 4000$---8000 K by 100 K step, 
$L_{\rm WD,0}= 0$---26,000 $L_\sun$ by 100 $L_\sun$ step,
and $i=70$---$90\arcdeg$ by $1\arcdeg$ step and seek
the best-fit model by the visual comparison of the model with the data.
After obtaining the best fit model for symmetric configurations,
we further seek the best fit model for August 2000 data 
by changing the asymmetric configuration parameter, 
$\zeta_{\rm low}$, from 1.0 to 0.05 by 0.05 step. 
The best fit asymmetric models are also shown in Figures 
\ref{vmag_quiescence_late_m10}-\ref{vmag_quiescence_late_m20}
only for August 2000 data.  Dashed lines indicate the symmetric
case while solid lines denote the asymmetric case.  
The adopted parameters are tabulated 
in Table \ref{system_parameters_CI_Aql}.  Enlargement around 
the eclipse is also shown in Figure \ref{vmag_eclipse_late_m15} 
only for the $1.5 M_\sun$ MS companion. 

\par
     We have calculated the absolute $M_B$,$M_V$, and $M_I$
magnitudes of our model \citep[see][in more detail]{hac01kb}.
The direct fit with the observational apparent magnitude, $m_V$,
gives us a distance modulus, $(m-M)_V$, as shown in Figures
\ref{vmag_quiescence_late_m10}-\ref{vmag_quiescence_late_m20}.
Since each direct fit determine a distance modulus, we have 
several different distance moduli for each light curve.
We have to choose the most appropriate distance modulus among them.
This value is iteratively sought in such a way that one distance 
modulus is consistent all among the light curves of the late phase
(May/April of 2001), the quiescent phase, the 1917 outburst, and 
the 2000 outburst as well as the color excess of $E(B-V)$, all of 
which will be discussed later in more detail.
\par
     The light curves are calculated for three companion masses of 
$M_{\rm MS}= 1.0~M_\odot$, $1.5~ M_\odot$ and $2.0~M_\odot$ as shown
in Figures \ref{vmag_quiescence_late_m10}-\ref{vmag_quiescence_late_m20}.
Since we have obtained similar light curves for these three cases,
we cannot determine the companion mass without additional information.
One may distinguish them from the original temperature
of the companion in quiescence; 
it is too high in the case of $1.0~M_\sun$ 
($7,500-7,700$~K vs. $T_{\rm ZAMS}= 5,600$~K) and too low in the case 
of $2.0~M_\sun$ ($6,900-6,700$~K vs. $T_{\rm ZAMS}= 9,100$~K)
compared with their zero-age main-sequence (ZAMS) surface temperature
\citep{bre93}.  Therefore, we adopt $M_{\rm MS}= 1.5 ~M_\odot$
in the following discussion.  
\par
     The mass transfer 
is as high as $\sim 1 \times 10^{-7} M_\sun$ yr$^{-1}$, and this
indicates a thermally unstable mass transfer \citep{heu92},
which requires $M_{\rm MS}/M_{\rm WD} \gtrsim 1.1$ \citep{web85}.
Moreover, of our best-fit model, the original surface temperature 
of the main-sequence companion is roughly consistent with
the effective temperature of a $\sim 1.5~M_\sun$ 
zero-age main-sequence star ($T_{\rm MS,org}= 7,300-7,100$~K vs.
$T_{\rm ZAMS}= 7200$~K).
Since the radius of the $1.5 ~M_\sun$ ZAMS star 
is about $1.45 ~R_\sun$, this 1.5 $M_\sun$ main-sequence companion 
has expanded to fill the Roche lobe of $\sim 1.7 ~R_\sun$
after the central hydrogen has decreased from $X=0.7$ 
at the zero-age to $X= 0.5$ at the age of $\sim 1$ G~yr \citep{bre93}. 
If the companion has already lost significant mass 
to the white dwarf component,
its original zero-age mass might be more massive than 1.5 $M_\sun$.
\par
\begin{figure}
\plotone{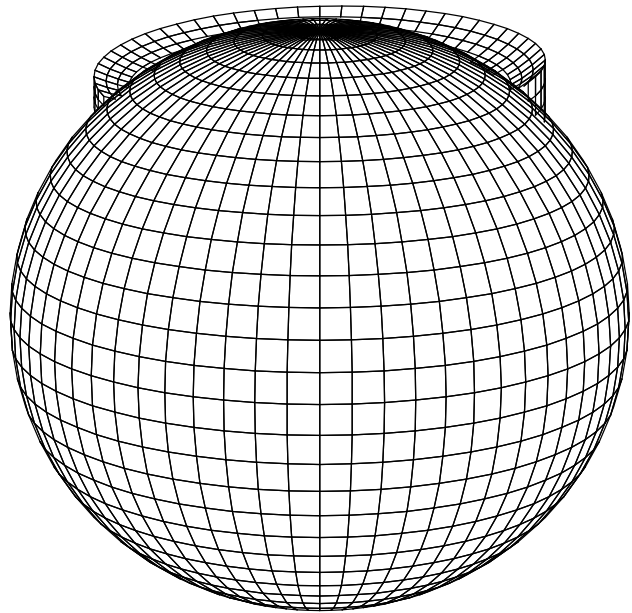}
\caption{
Same as those in Fig.\ref{ciaql_late_config}, but for 
the configuration at the eclipse minimum.
\label{ciaql_late_config_eclipse}}
\end{figure}

     The edge of the accretion disk is not fully occulted by the
companion star at eclipse minima as shown 
in Figure \ref{ciaql_late_config_eclipse}.  This partial occultation
implies a shift of the phase position at eclipse minima.
It may advance when the asymmetry of the accretion disk 
is relatively large and the inner side of the spray is
brightly illuminated by the white dwarf.  However, our best fit
model in Figure \ref{vmag_eclipse_late_m15} shows no significant
advance in the orbital phase at eclipse minima mainly because
the asymmetry of $\zeta_{\rm low}=0.7$ is not so large.   
\par
     We are able to estimate the turn-off time of steady hydrogen
shell-burning on the white dwarf.  The orbital light curve fitting
for May 2001 suggests us that the intrinsic luminosity of 
the white dwarf has already decreased to 
about $L_{\rm WD,0} \sim 11,000 ~L_\sun$ from $26,000~L_\sun$ 
for the lower limit of steady hydrogen shell-burning.
The orbital light curve fitting for August of 2001 
indicates that the intrinsic luminosity of the white dwarf
has much decreased to about $L_{\rm WD,0} \sim 2,200~L_\sun$. 
Thus, we have estimated the turn-off time a bit before May of 2001,
that is, March or April of 2001.
After the turn-off time, the white dwarf began to cool down.
It should be noted, however, that hydrogen shell burning does not 
vanish suddenly, but gradually extinguishes.  Then,
our theoretical definition of the turn-off time is when 
the nuclear luminosity decreases below the diffusive luminosity
of the white dwarf.  In other words, the diffusive luminosity 
of the white dwarf is almost balanced with the nuclear luminosity 
before the turn-off time.  Nuclear burning still partly contributes
to the luminosity of the white dwarf even after the turn-off time. 
A large part of the luminosity in the cooling phase comes from 
the thermal energy in the hydrogen-rich envelope and the helium 
shell (ash of hydrogen burning) of the white dwarf. 
\par
\begin{figure}
\plotone{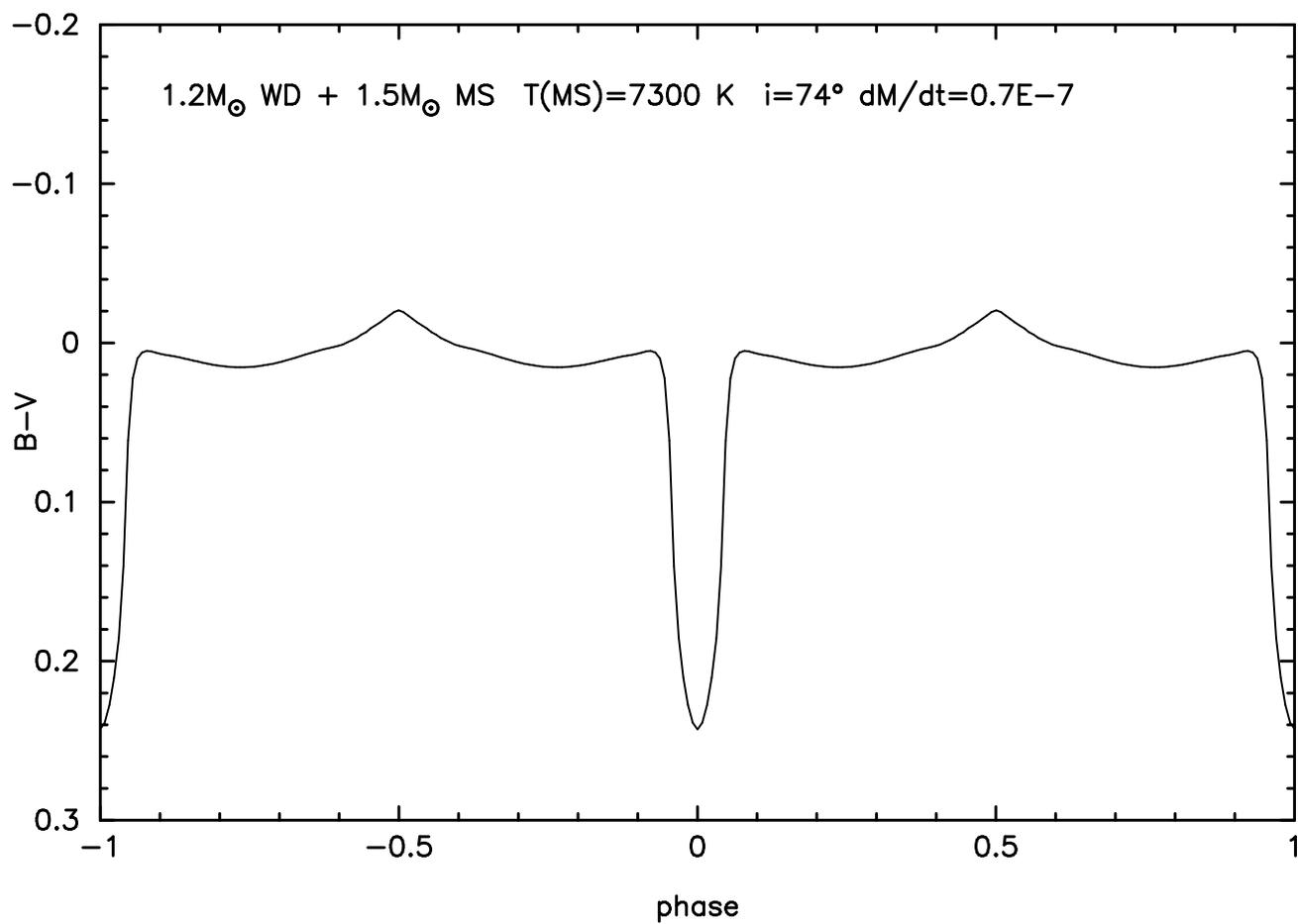}
\caption{
Calculated $B-V$ light curve in quiescence is plotted 
against the orbital phase for a pair of $1.2~M_\sun$ WD and
$1.5~M_\sun$ MS companion.
\label{color_quiescence_late}}
\end{figure}

     We have calculated the color index 
to estimate the color excess of CI Aql.
A rather blue color index of $(B-V)_{\rm c} \approx 0.0$ 
is obtained for the case of a $1.5~M_\sun$ companion
as shown in Figure \ref{color_quiescence_late}.  
This suggests a large color excess of 
$E(B-V)= (B-V)_{\rm o} - (B-V)_{\rm c} \approx 1.0$
with the observed color of $(B-V)_{\rm o} \sim 1.0$ 
\citep[see Table 2 of][]{men95}, where $(B-V)_{\rm c}$ is the calculated
color and $(B-V)_{\rm o}$ is the observed color. 
This color excess is roughly consistent with $E(B-V) = 0.85 \pm 0.3$ 
estimated by \citet{kis01}.  Therefore, we adopt $E(B-V) = 1.0$ 
in this paper.

\placefigure{vmag_quiescence_late_m10}
\placefigure{vmag_quiescence_late_m15}
\placefigure{vmag_quiescence_late_m20}
\placefigure{vmag_eclipse_late_m15}
\placefigure{ciaql_late_config_eclipse}
\placefigure{color_quiescence_late}
\placetable{system_parameters_CI_Aql}

\section{THE 1917 OUTBURST}
\begin{figure}
\plotone{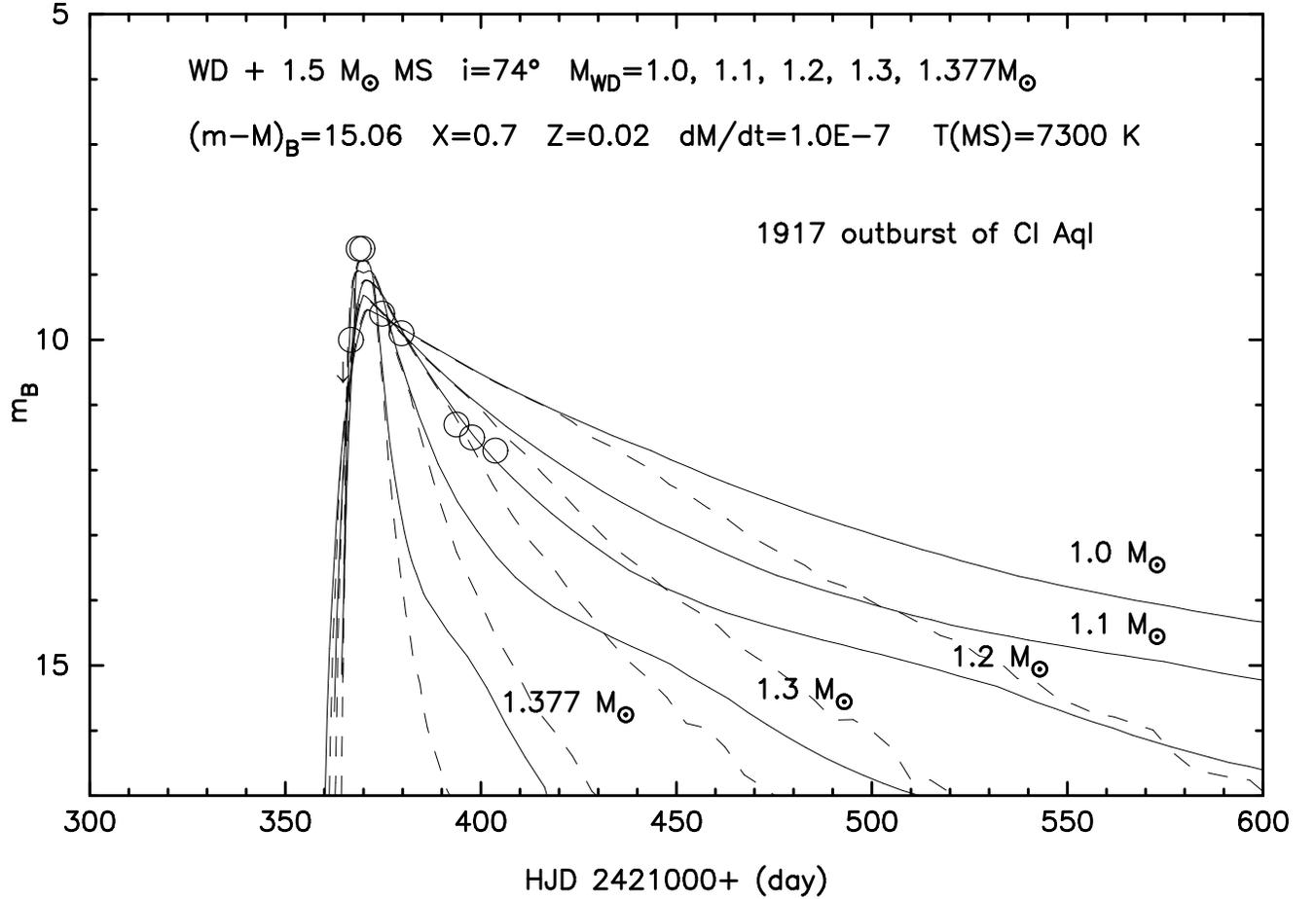}
\caption{
Calculated $B$ light curves are plotted against time (HJD 2,421,000+) 
together with the observational points of the 1917 outburst.
The WD mass is attached to each light curve.
Open circles indicate observational points 
\citep[taken from][]{wil00db}.
The model consists of a bloated white dwarf 
photosphere with no accretion disk
and a nonirradiated main-sequence companion. 
The hydrogen content of the white dwarf envelope 
is assumed to be $X=0.70$ for all models.  
The apparent distance modulus of $(m-M)_B= 15.06$ is assumed
for all the white dwarf masses.  
Solid lines indicate the light curves connecting the $B$ light 
at the binary phase 0.5 (roughly the mean brightness)
while dashed lines correspond 
to those connecting the $B$ light at the binary phase 0.0
(mid eclipse).
\label{bmag_mmix_ciaql1917}}
\end{figure}

     The light curve of the 1917 outburst has also been re-analyzed
with the newly obtained system parameters.  
The Tycho $B$ magnitude of the 1917 outburst has been reported 
by \citet{wil00db}, which is shown in Figure \ref{bmag_mmix_ciaql1917}
together with our calculated $B$ light curves.
We have obtained the apparent distance modulus of $(m-M)_B= 15.06$ 
based on the color excess of $E(B-V)= 1.0$ and 
the $V$-magnitude distance modulus estimation of $(m-M)_V= 14.06$. 
The numerical method for obtaining light curves is the same as
that in our previous one \citep{hac01ka} and it is fully described 
in \citet{hac01kb}.
\par
     Assuming a binary system consisting of a white dwarf with 
the mass of $M_{\rm WD}= 1.0$, 1.1, 1.15, 1.2, 1.3, and $1.377~M_\sun$
and a $1.5~M_\sun$ lobe-filling, main-sequence companion, 
we have calculated the developing of the outburst and obtained 
$B$ light curves for the hydrogen content of $X=0.7$.  
In Figure \ref{bmag_mmix_ciaql1917},
we do not include the effect of the accretion 
disk or the irradiation effect of the companion. 
These irradiation effects do not contribute much to the $B$ light
until the $B$-magnitude decreases to $B \sim 12$ mag, since the white
dwarf photosphere has expanded larger than the binary orbit \citep{kat94h}.
It is clear that there are no mid plateau phases around $V \sim 13.5$ 
for the case of no accretion disk and no irradiation of the companion.
\par 
     The white dwarf mass is again determined to be 
$M_{\rm WD}= 1.2 \pm 0.05 ~M_\sun$ 
from the fitting as seen in Figure \ref{bmag_mmix_ciaql1917}.
The various physical parameters of the outbursts are summarized 
in Table \ref{decline_rates}.
Here, $Z$ is the content of heavy elements by weight,
$\Delta M_{\rm max}$ is the hydrogen-rich envelope mass 
at the optical maximum, $\eta_{\rm wind}$ is the ratio
of the mass lost by winds, $\Delta M_{\rm wind}$, to $\Delta M_{\rm max}$,
i.e., $\eta_{\rm wind}= \Delta M_{\rm wind} / \Delta M_{\rm max}$,
$t_2$ is the time taken to drop 2 mag from maximum,
$t_3$ is the time to drop 3 mag from maximum,
$t_{\rm wind}$ is the duration of optically thick wind phase,
and $t_{\rm H}$ is the duration of steady hydrogen shell-burning.
\par
     We are not able to determine the hydrogen content 
only from the fitting with the 1917 light curve, because
much lower contents of hydrogen such as $X=0.5$ and $X=0.35$ 
by weight give a similar decline rate ($t_2$ or $t_3$) 
of the early phase light curve as listed in Table \ref{decline_rates}.

\placefigure{bmag_mmix_ciaql1917}
\placetable{decline_rates}

\section{THE 2000 OUTBURST}
     CI Aql erupted in 2000 April and was densely observed
in various optical bands \citep[e.g.,][]{kis01,mat01}.
We have determined the hydrogen content of the white dwarf envelope
from the turn-off time of the 2000 outburst.
\par
     We summarize the global feature of the optical light curve:
its optical maximum ($V \sim 9$ mag) was reached on 2000 May 5
\citep[HJD 2451669.5,][]{kis01}.
As shown in Figure \ref{vmag_irradmix_ciaql00_m15},
the visual brightness quickly decreased to 13.5 mag 
in about 60 days.  Then, it stays at $V \sim 14$ mag, i.e.,
a plateau phase.
This mid plateau phase is very similar 
to that of U Sco and can also be explained by the irradiation 
of the accretion disk \citep{hkkm00, hkkmn00, tho01}.
Therefore, we have reproduced the light curve in the plateau phase
by assuming model parameters similar to those of U Sco.
\par
     Theoretical light curves in Figure \ref{vmag_irradmix_ciaql00_m15}
are calculated for a pair of a $1.2 ~M_\odot$ white dwarf
and a $1.5 ~M_\odot$ main-sequence companion.
In the plateau phase, the light curve is determined mainly by 
the irradiations of the accretion disk and the main-sequence companion,
because the white dwarf photosphere becomes 
much smaller than the binary size.  
The unheated surface temperatures of the companion are fixed 
to be $T_{\rm MS, org}= 7,300$ K throughout the 2000 outburst
and $T_{\rm disk, rim}= 6,600$ K at the disk rim, which are
the same as the model parameters in the late phase of the 2000 
outburst.
\par
     The luminosity of the accretion disk depends on 
both the thickness $\beta$ and the size $\alpha$.
Here, we change $\alpha$ by 0.1 step and $\beta$ by 0.05 step
and seek the best fit model for the wind phase, 
the steady hydrogen shell-burning phase, and the cooling phase.
Finally we have adopted a set of $\alpha=3.0$ and $\beta=0.05$
during the wind phase to reproduce the orbital modulation observed
by \citet{mat01}, $\alpha=0.8$ 
and $\beta=0.30$ in the steady hydrogen shell-burning phase,
and $\alpha=0.8$ and $\beta=0.20$ in the cooling phase.
Here, we adopt symmetric configurations of the accretion disk 
throughout the 2000 outburst.
\par
     The decline of the early phase ($t \sim 0$---70 days) 
hardly depends on the hydrogen content, $X$, of the white dwarf envelope, 
but the ends of the wind phase and of the hydrogen shell-burning 
phase depend sensitively on the hydrogen content as tabulated
in Table \ref{decline_rates}.
Therefore, we have finally determined the hydrogen content of $X=0.35$ 
and shown calculated light curves 
in Figure \ref{vmag_irradmix_ciaql00_m15} for three different bands
of $B$, $V$, and $I_c$. 
\par
     Here, we have used the apparent distance modulus of
$(m-M)_V= 14.06$ for the $V$ band.  Adopting the color excess
of $E(B-V)= 1.0$, we obtain the apparent distance moduli
of $(m-M)_B= 15.06$ and $(m-M)_I= 12.46$ for $B$ and $I_c$ 
bands, respectively.   Here, we adopted the relation
of $E(I-V) / E(B-V)= -1.60$ \citep[e.g.,][]{rie85}.
Both the $B$ and $I_c$ light curves are in good 
agreement with the observational points as shown in
Figure \ref{vmag_irradmix_ciaql00_m15}.  
This indicates the consistency of our estimated values.
Various physical values of the 2000 outburst are summarized 
in Table \ref{envelope_mass2000}.
\par
     \citet{mat01} reported a sharp $\sim 1$ mag drop 
of $R_c$-magnitude on 23 November 2000 and Kiyota 
(2001, VSNET archives, http://vsnet.kusastro.kyoto-u.ac.jp/vsnet/)
also observed a $\sim 1.5$ mag drop of $I_c$-magnitude
around the same day as shown in Figure \ref{vmag_irradmix_ciaql00_m15}.
If we attribute this drop to the end of wind phase,
the hydrogen content of $X=0.35$ is consistent with the drop.
\par
     We have re-estimated the luminous supersoft X-ray source phase. 
No luminous supersoft X-rays are expected in the early decline phase
because the photospheric radius of the white dwarf is rather large 
and the photospheric temperature is relatively low.
After the recurrent nova enters the plateau phase, 
the photospheric temperature of the white dwarf increases
enough to emit supersoft X-rays \citep{kat99}.  
During the massive wind phase, however, we do not expect
supersoft X-rays because they are self-absorbed by the wind itself. 
U Sco was observed as a luminous supersoft X-ray source
in the mid plateau phase of the 1999 outburst just after
the massive wind stopped \citep{kah99,hkkm00}.
Our analysis suggests that the massive wind stopped before
23 November 2000.  
This means that a luminous supersoft X-ray phase started
from November of 2000 and continued until March of 2001 as shown
in Figure \ref{vmag_irradmix_ciaql00_m15}.  The rather low X-ray fluxes
in June and August of 2001 \citep{gre02}
are also very consistent with our analysis.

\placefigure{vmag_irradmix_ciaql00_m15}
\placetable{envelope_mass2000}

\section{DISCUSSION}
\begin{figure}
\plotone{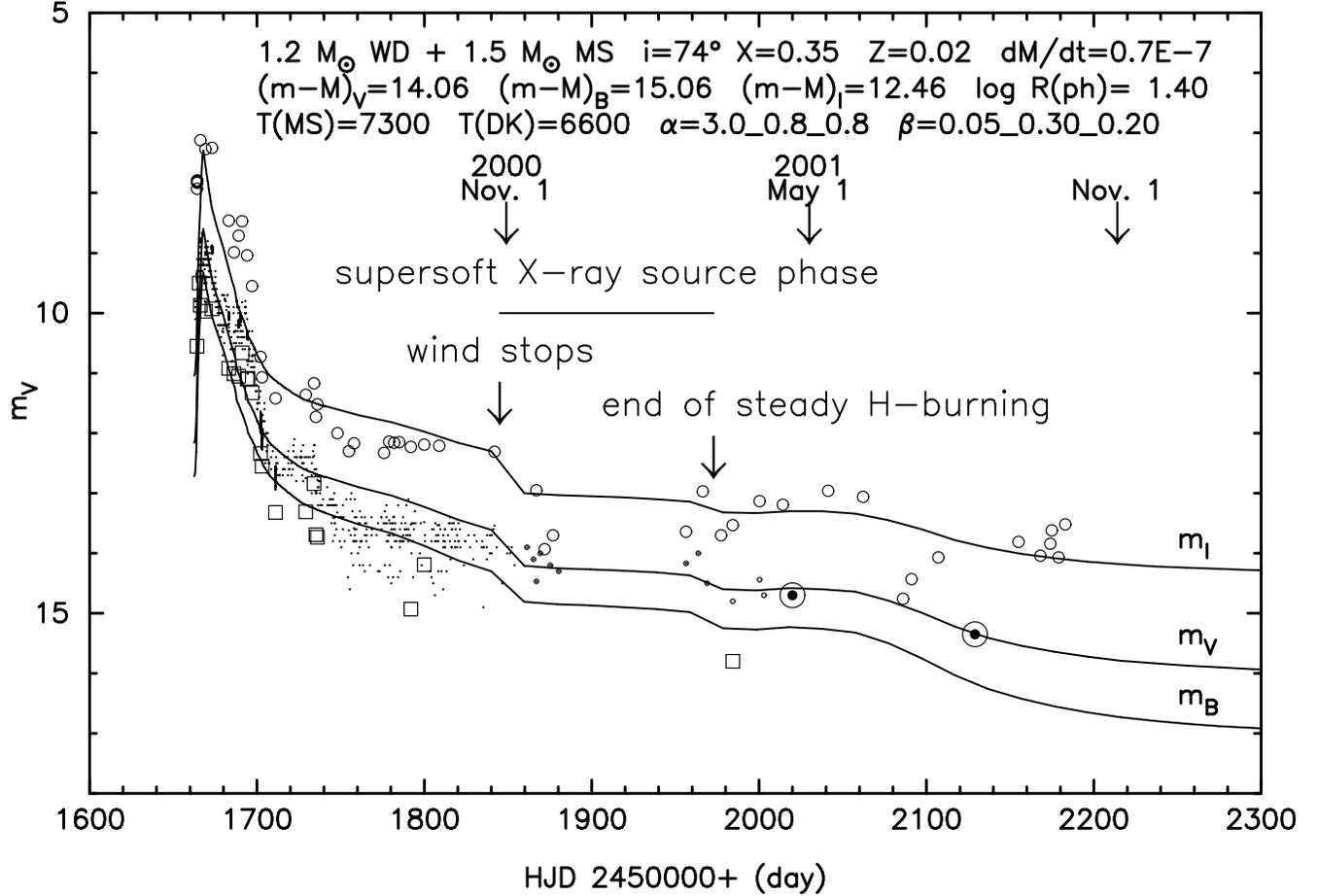}
\caption{
Calculated $V$, $B$,  and $I_c$ light curves are plotted 
against time (HJD 2,450,000+) 
together with the observational points of the CI Aql 2000 outburst.
Small dots (small filled circles, small open circles) 
indicate observational $V$ and visual magnitudes, 
while open squares represent observational $B$ magnitudes
and large open circles indicate observational $I_c$ magnitudes 
(all taken from VSNET archives).  Two $V$-magnitudes in the late-phase 
observed by \citet{sch01a, sch01b} are indicated by $\sun$ marks.  
Each light curve connects the brightness at the binary phase 0.4.  
The apparent distance moduli are $(m-M)_V= 14.06$, $(m-M)_B= 15.06$
and $(m-M)_I= 12.46$ for the $V$, $B$,  and $I_c$ light curves,
respectively.
\label{vmag_irradmix_ciaql00_m15}}
\end{figure}

     The apparent distance modulus is obtained to be $(m-M)_V= 14.06$
by fitting for the late phase of the 2000 outburst. 
The color excess is also estimated to be $E(B-V)=1.0$ 
from the difference between the observed color and the calculated 
color, and the absorption of $A_V=3.1$ 
from a relation of $A_V = 3.1~E(B-V)$ \citep[e.g.,][]{rie85}.  
Therefore, the distance to CI Aql is derived to be 1.55 kpc.
The distance modulus of $(m-M)_B= 15.06$ is derived from 
the distance modulus of $(m-M)_V= 14.06$ and
the color excess of $E(B-V)=1.0$, which are also consistent with 
the $B$ light curve of the 1917 outburst.  
This set of $(m-M)_V= 14.06$ and
$E(B-V)=1.0$ are also in very good agreement with the $V$, $B$,
and $I_c$ light curves of the 2000 outburst 
as shown in Figure \ref{vmag_irradmix_ciaql00_m15}.
\par
\begin{figure}
\plotone{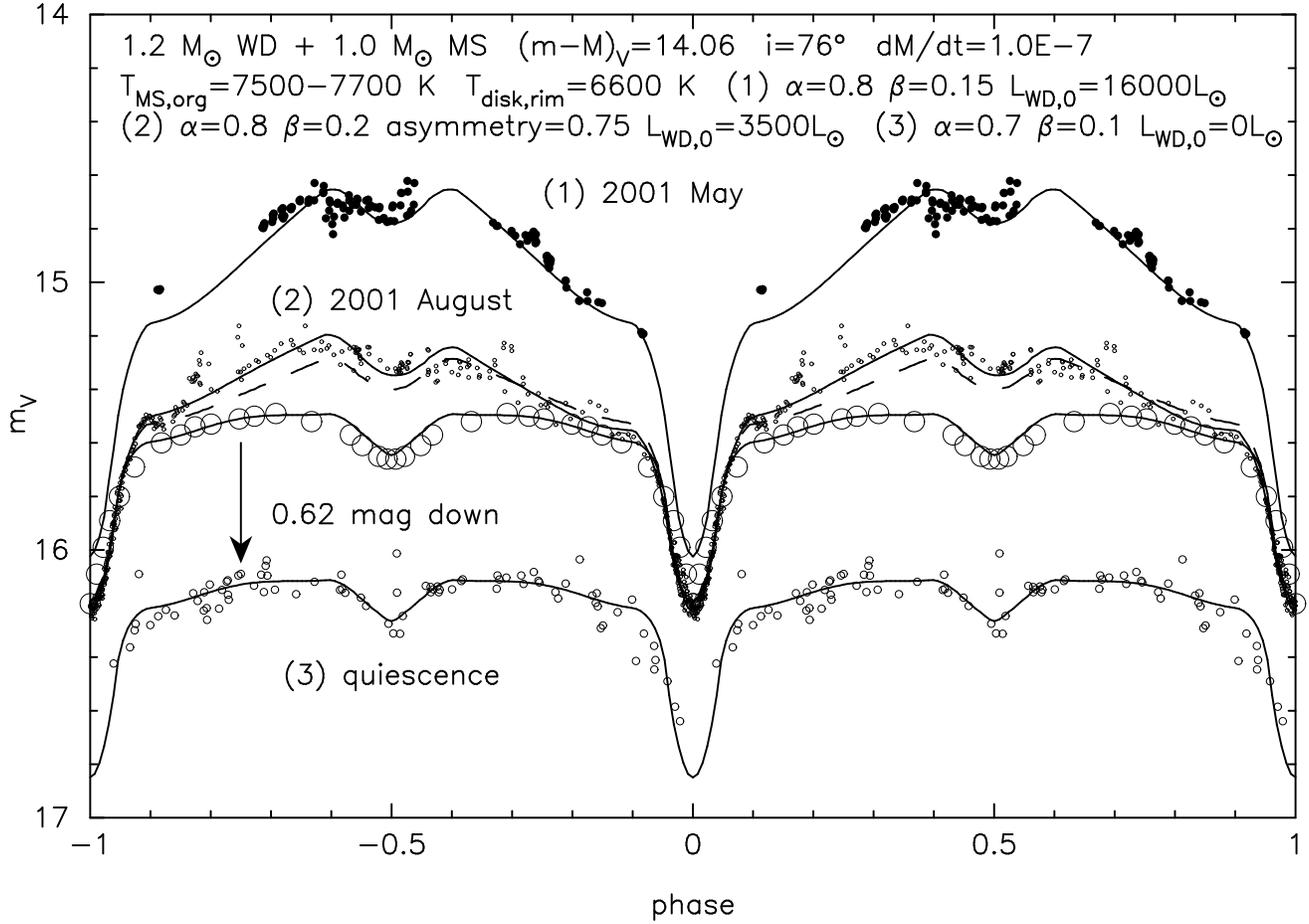}
\caption{
Calculated $V$ light curves are plotted against the binary phase 
(binary phase is repeated twice from $-1.0$ to $1.0$)
together with the observational points of the two late phases 
(filled circles are in April/May of 2001, 
and small open circles are in August of 2001), and
the quiescent phase (large open circles: smoothly averaged by us 
from \citet{men95}) as well as the revised quiescent phase data
(middle size open circles: Honeycutt 2002, private communication).
The visual magnitude in quiescence has been revised by 0.62 mag
down as seen in the figure.
Here we adopted the ephemeris given by \cite{mat01}
as in equation (\ref{new_ephemeris}).  
The model is a binary system of $1.2 M_\odot$ white dwarf (WD) 
$+$ $1.0 M_\odot$ main-sequence (MS) companion. 
The other model parameters are shown in the figure and Table 
\ref{system_parameters_CI_Aql}. 
The solid line for August of 2001 indicates the case of
asymmetric structure of accretion disk while the dashed
line denotes the case of symmetric configuration of accretion disk.
The other solid lines correspond to the symmetric accretion disks. 
Here ``asymmetry=0.75'' means the value of $\zeta_{\rm low}=0.75$.
\label{vmag_quiescence_late_m10}}
\end{figure}
\begin{figure}
\plotone{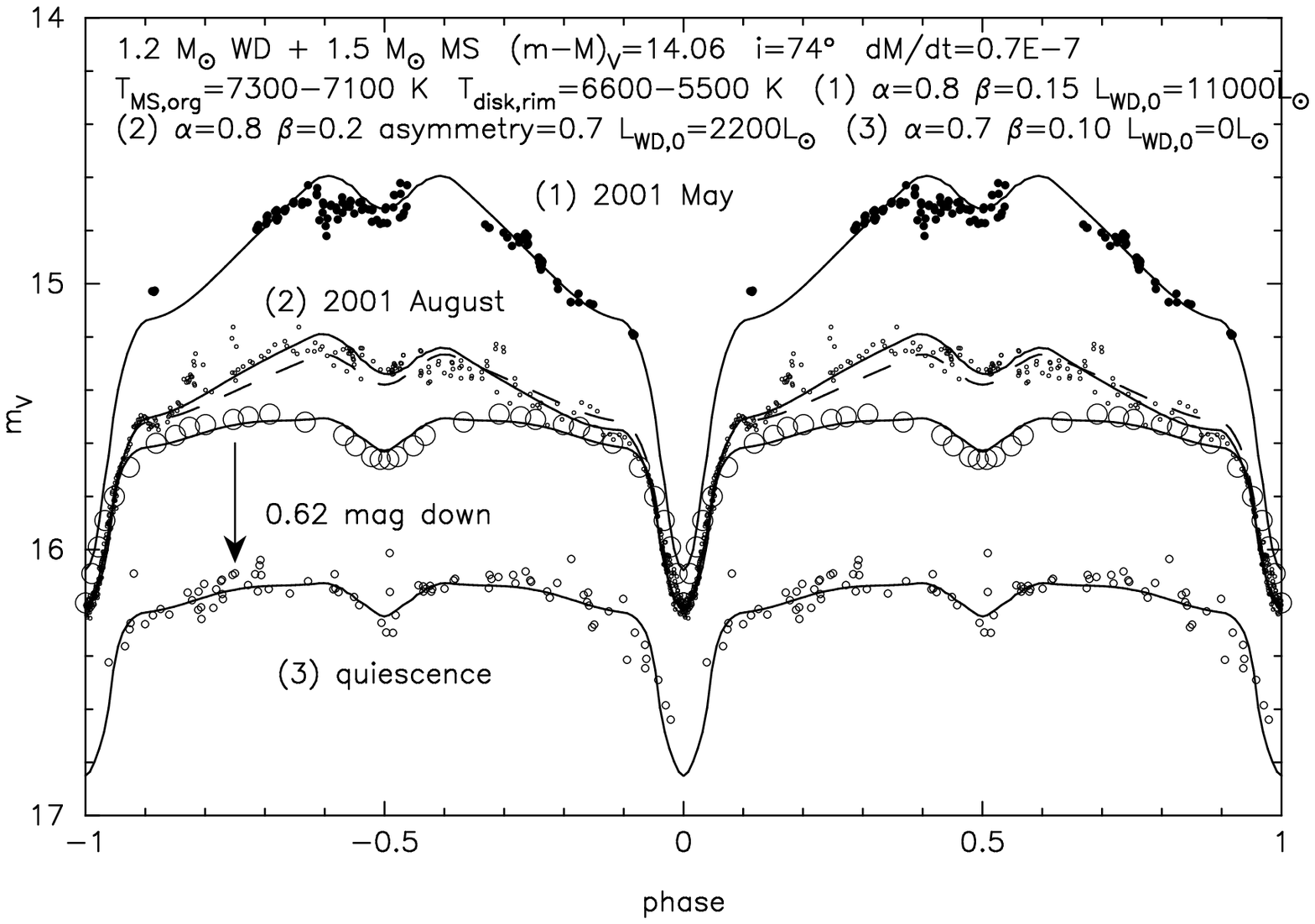}
\caption{
Same as those in Fig.\ref{vmag_quiescence_late_m10}, but for 
a binary system of $1.2 M_\odot$ WD $+$ $1.5 M_\odot$ MS companion. 
\label{vmag_quiescence_late_m15}}
\end{figure}
\begin{figure}
\plotone{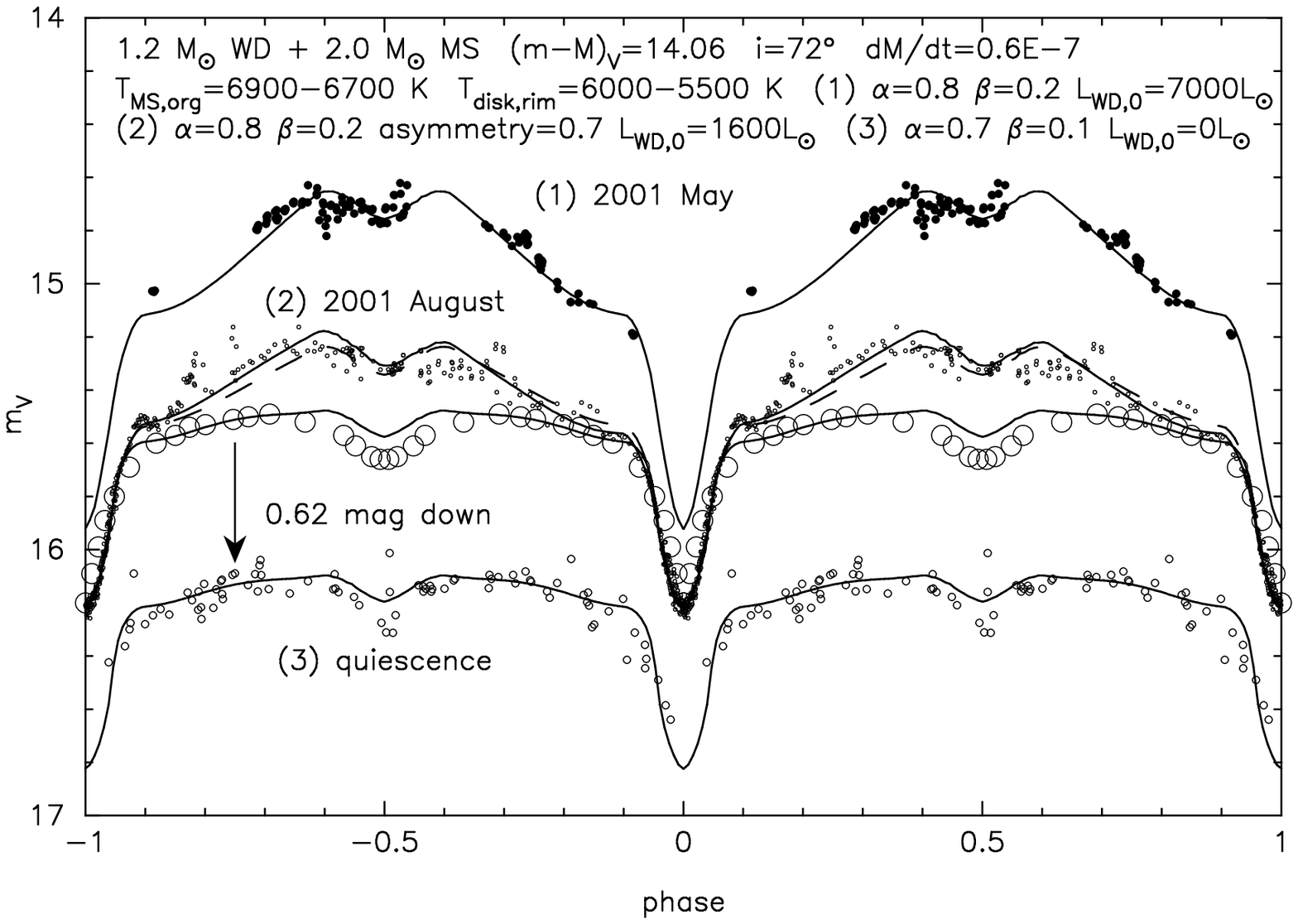}
\caption{
Same as those in Fig.\ref{vmag_quiescence_late_m10}, but for 
a binary system of $1.2~M_\odot$ WD $+$ $2.0~M_\odot$ MS companion. 
\label{vmag_quiescence_late_m20}}
\end{figure}
\begin{figure}
\plotone{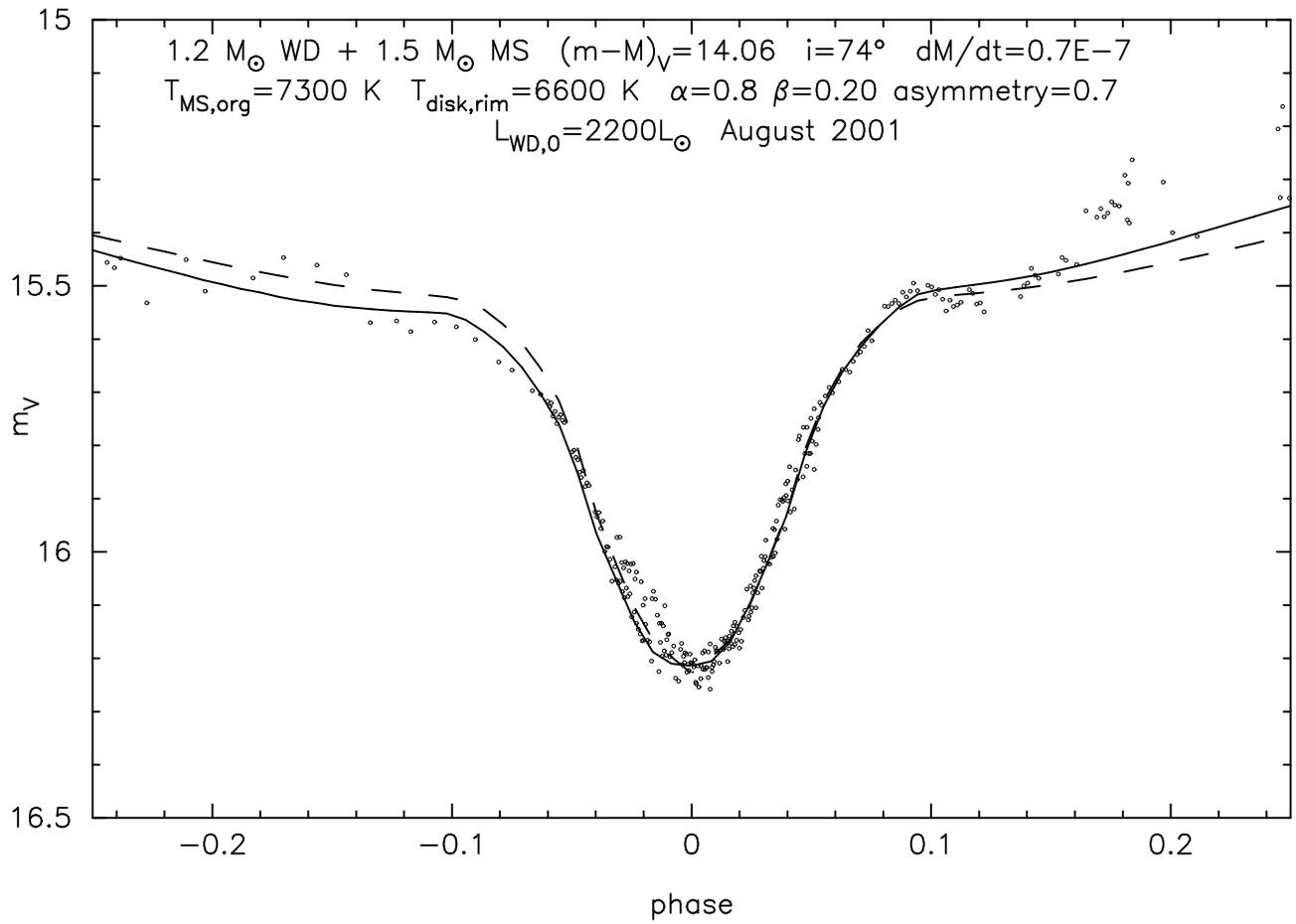}
\caption{
Same as those in Fig.\ref{vmag_quiescence_late_m15}, but for 
an enlargement around eclipse minima.
Only the data of August 2001 are shown together with
the model light curves, solid line for asymmetric and dashed one
for symmetric configurations of the accretion disk. 
\label{vmag_eclipse_late_m15}}
\end{figure}

     The orbital light curve in August of 2001 shows a significant
asymmetry.  This indicates that the shape of the accretion disk
is changing in time at least in the final decay phase of the 2000 
outburst.
Assuming the asymmetry of $\zeta_{\rm low}= 0.7$ or $0.75$, 
we have calculated a $V$ light curve and added it to Figures
\ref{vmag_quiescence_late_m10}-\ref{vmag_quiescence_late_m20}
(solid line for August of 2001).  On the other hand, the disk shape 
is almost symmetric for May of 2001 and in quiescence. 
It should be noted that the model light curve of August 2001 
does not reproduce the observational lack of any secondary 
eclipses even if we introduce asymmetry of the accretion disk shape.
It may stem from large time variations of the edge of the accretion 
disk.  
\par
     The envelope mass at the visual maximum is estimated
to be $\Delta M_{\rm max}= 8.0 \times 10^{-6} M_\sun$ for
the hydrogen content of $X=0.35$, indicating the average 
mass transfer rate of $\dot M_{\rm acc}= 1.0 \times 10^{-7} 
M_\sun$ yr$^{-1}$ between the 1917 and 2000 outbursts, or
$\dot M_{\rm acc}= 1.3 \times 10^{-7} 
M_\sun$ yr$^{-1}$ between the 1941 and 2000 outbursts.
This mass accretion rate is a bit higher but almost consistent with 
our CI Aql model in quiescence (see Figs.
\ref{vmag_quiescence_late_m10}-\ref{vmag_quiescence_late_m20}).
\par
     On the other hand,
if the mean recurrence period is 20 yrs as suggested 
by \citet{sch01c}, the average mass transfer rate increases
to $\dot M_{\rm acc} \sim 4 \times 10^{-7} M_\sun$ yr$^{-1}$.
This value is high enough to maintain steady hydrogen shell-burning
for hydrogen content of $X=0.7$ on a $1.2~M_\sun$ WD when the mass
transfer is steady.  In other words, it is too high to be compatible
with our thermonuclear runaway model, unless the mass transfer rate
itself rapidly increased just before the outburst.  
When the hydrogen content of the transferred matter is
as low as $X \sim 0.35$, the lower limit to steady hydrogen 
shell-burning becomes $8 \times 10^{-7} M_\sun$~yr$^{-1}$
\citep[see e.g. eq.(7) in][]{hac01kb}.  Then, the mass accretion rate of 
$\dot M_{\rm acc} \sim 4 \times 10^{-7} M_\sun$ yr$^{-1}$
can still cause shell-flashes on a $1.2~M_\sun$ WD, although 
the orbital modulation of the light curve in quiescence is not 
consistent with such a high mass transfer rate as shown in Figures
\ref{vmag_quiescence_late_m10}-\ref{vmag_quiescence_late_m20}.
\par
     About 82\% of the envelope mass is lost by the wind 
($\Delta M_{\rm wind}= 6.6 \times 10^{-6} M_\sun$),
while the residual 18\% ($\Delta M_{\rm He}= 1.4 \times 10^{-6} M_\sun$)
is left and added to the helium layer.  The net mass increasing
rate of the white dwarf is about 
$\dot M_{\rm He}= 1.7 \times 10^{-8} M_\sun$ yr$^{-1}$
between the 1917 and 2000 outburst or
$\dot M_{\rm He}= 2.4 \times 10^{-8} M_\sun$ yr$^{-1}$
between the 1941 and 2000 outburst.
These satisfy the conditions of Type Ia supernova explosion
if the white dwarf consists of carbon and oxygen
\citep{nom91}.  Evolutionary paths to Type Ia supernovae via
recurrent novae have been discussed in more detail in
\citet{hkn99, hknu99}.



\acknowledgments
     We thank the VSNET members who observed CI Aql
and R. K. Honeycutt for his providing us the unpublished data 
of CI Aql in quiescence.
This research has been supported in part by the Grant-in-Aid for
Scientific Research (11640226) 
of the Japan Society for the Promotion of Science.

\begin{deluxetable}{lllll}
\tabletypesize{\scriptsize}
\tablecaption{Fitted system parameters of CI Aquilae\tablenotemark{a} 
\label{system_parameters_CI_Aql}}
\tablewidth{0pt}
\tablehead{
\colhead{parameter} & 
\colhead{symbol} & 
\colhead{$M_2=1.0~M_\odot$} &  
\colhead{$M_2=1.5~M_\odot$} & 
\colhead{$M_2=2.0~M_\odot$}  
} 
\startdata
inclination angle & $i$ & $76 \arcdeg$ & $74 \arcdeg$ & $72 \arcdeg$ \\
WD luminosity (May of 2001) & $L_{\rm WD,0}$ & 16,000 $L_\sun$ & 11,000 $L_\sun$ & 7,000 $L_\sun$ \\
WD luminosity (August of 2001) & $L_{\rm WD,0}$ & 3,500 $L_\sun$ & 2,200 $L_\sun$ & 1,600 $L_\sun$ \\
MS temperature (May/August of 2001) & $T_{\rm MS,org}$ & 7,500 K & 7,300 K & 6,900 K \\
MS temperature (quiescence) & $T_{\rm MS,org}$ & 7,700 K & 7,100 K & 6,700 K \\
color excess & $E(B-V)$ & \nodata & 1.0 & \nodata \\
distance modulus & $(m-M)_V$ & \nodata & 14.06 & \nodata \\
absorption & $A_V$ & \nodata & 3.1 & \nodata \\
distance & $d$ & \nodata & 1.55 kpc & \nodata 
\enddata
\tablenotetext{a}{The white dwarf mass is assumed to 
be $M_{\rm WD}= 1.2 M_\odot$ 
}
\end{deluxetable}

\begin{deluxetable}{llllrrrr}
\tabletypesize{\scriptsize}
\tablecaption{Theoretical decline rates of light curves 
for the CI Aql 1917 outburst  
\label{decline_rates}}
\tablewidth{0pt}
\tablehead{
\colhead{$M_{\rm WD}$} & 
\colhead{$X$\tablenotemark{a}} & 
\colhead{$\eta_{\rm wind}$\tablenotemark{b}} &
\colhead{$t_2$} &
\colhead{$t_3$} &
\colhead{$t_{\rm wind}$} &
\colhead{$t_{\rm H}$} 
\\
\colhead{$(M_\sun)$} &
\colhead{} &
\colhead{} &
\colhead{(day)} &
\colhead{(day)} &
\colhead{(day)} &
\colhead{(day)} 
} 
\startdata
1.377 & 0.7 & 0.95 & 3.5 & 5 & 49 & 59 \\
1.3 & 0.7 & 0.90 & 9 & 14 & 148 & 287 \\
1.2 & 0.35 & 0.82 & 20 & 35 & 186 & 310 \\
1.2 & 0.5 & 0.83 & 21 & 37 & 219 & 450 \\
1.2 & 0.7 & 0.84 & 22 & 39 & 266 & 720 \\
1.15 & 0.7 & 0.82 & 28 & 48 & 330 & 1020 \\
1.1 & 0.7 & 0.82 & 34 & 61 & 429 & 1450 \\
1.0 & 0.7 & 0.82 & 55 & 78 & 730 & 1540 
\enddata
\tablenotetext{a}{Solar metallicity is assumed, i.e., $Z=0.02$.
}
\tablenotetext{b}{The ratio of the mass lost by winds 
to the accretion mass, i.e.,
$\Delta M_{\rm wind} / \Delta M_{\rm max}$.
}
\end{deluxetable}

\begin{deluxetable}{lllllllcccc}
\tabletypesize{\scriptsize}
\tablecaption{Theoretical properties of the CI Aql 2000 outburst  
\label{envelope_mass2000}}
\tablewidth{0pt}
\tablehead{
\colhead{$M_{\rm WD}$} & 
\colhead{$X$} & 
\colhead{$Z$\tablenotemark{a}} & 
\colhead{$\Delta M_{\rm max}$} &
\colhead{$\Delta M_{\rm wind}$} &
\colhead{$\eta_{\rm wind}$} & 
\colhead{$\Delta M_{\rm He}$} &
\colhead{$t_2$} &
\colhead{$t_3$} &
\colhead{$t_{\rm wind}$} &
\colhead{$t_{\rm H}$} 
\\
\colhead{$(M_\sun)$} &
\colhead{} &
\colhead{} &
\colhead{$(M_\sun)$} &
\colhead{($M_\sun$)} &
\colhead{} &
\colhead{($M_\sun$)} &
\colhead{(day)} &
\colhead{(day)} &
\colhead{(day)} &
\colhead{(day)} 
} 
\startdata
1.2 & 0.35 & 0.02 & $8.0 \times 10^{-6}$ & $6.6 \times 10^{-6}$ & 0.82 
& $1.4 \times 10^{-6}$ & 18 & 30 & 180 & 307 
\enddata
\tablenotetext{a}{Solar metallicity is assumed}
\end{deluxetable}






\clearpage

\clearpage

\clearpage

\clearpage

\clearpage

\clearpage

\clearpage

\clearpage


\end{document}